\documentclass[preprint,aps,prl,showpacs]{revtex4}
\usepackage{epsfig,pstricks}
\usepackage{amsmath,amssymb}

\begin{document}

\title{Lava channel formation during the 2001 eruption on Mount Etna: 
       evidence for mechanical erosion}
\author{Carmelo Ferlito$^1$}
\author{Jens Siewert$^{2}$}
\affiliation{$^1$ Dipartimento Scienze Geologiche, Universit\`a di Catania, 
                  I-95125 Catania, Italy\\
            $^2$ MATIS-INFM $\&$ Dipartimento di Metodologie Fisiche e
                 Chimiche (DMFCI), viale A. Doria 6, 95125 Catania, Italy}

\begin{abstract}
We report  the direct observation of a peculiar lava channel that was 
formed near the base of a parasitic cone during the 2001 eruption on 
Mount Etna. Erosive processes by flowing lava are commonly attributed 
to thermal erosion. However, field evidence strongly suggests that models 
of thermal erosion cannot explain the formation of this channel. 
Here, we put forward the idea that the essential erosion mechanism 
was abrasive wear. By applying a simple model from tribology 
we demonstrate that the available data agree favorably with our hypothesis. 
Consequently, we propose that erosional processes resembling 
the wear phenomena in glacial erosion are possible in a volcanic environment.
\end{abstract}


\maketitle

{\em Introduction} -- 
Formation of lava channels is usually discussed in terms of thermal or 
thermomechanical erosion~\cite{1,2,3,7,7.1,8,9}. The basic idea is that heat 
is transferred from the hot flowing lava to the cold substrate, 
thereby melting it and washing it out. Thermal erosion by lava 
has been studied quantitatively both for turbulent~\cite{1,3,7,7.1} and for 
laminar flows~\cite{2,8,9}. 
The physics underlying the quantitative description is that of heat 
transport in liquids and solids combined with fluid dynamics. 
The idea of  thermal erosion in the formation of lava channels is 
supported by field evidence only in few examples~\cite{11,12,13,14,10}, e.g., 
in archaean komatiite lavas~\cite{11}, in African carbonatites~\cite{12,13}, 
or in basaltic lava tubes of Kilauea volcano (Hawaii)~\cite{14}.

In the picture of thermal erosion, the relevance of mechanical 
processes appears to be secondary being mainly limited to the 
removal of the melted material. However, it has been noted that 
mechanical effects may enhance the erosion rates~\cite{10,8}.  
Very recently, Williams {\it et al.} have reported geochemical evidence
for mechanical erosion in the Cave Basalt of Mount St.\ 
Helens where a basaltic flow eroded an unconsolidated
pyroclastic substrate~\cite{14.1}.  To our knowledge, this is the only example 
in the literature 
in which a dominant role of mechanical erosion of lava has been 
inferred (aspects of models for mechanical 
erosion have been considered theoretically before~\cite{15}). 
Therefore, the direct observation of the formation of a lava channel 
whose origin cannot be explained by  thermal erosion is of considerable 
interest. Such an observation may help to develop new and more complete 
models for the explanation of the field data elsewhere. Moreover, it 
generates interesting cross-links to other branches of geology where a 
variety of erosion models are discussed.

The starting point of this work was the formation of an uncommon channel 
during the 2001 eruption on Mount Etna which provides field evidence 
for purely mechanical erosion of lava flow. 
First, we briefly describe the morphological details of the channel and 
the history of the eruption (see also Ref.~\cite{15.1}). 
An estimate of the heat transfer required 
for the formation of the channel basically rules out thermal erosion as 
the essential process. Then, we will discuss the possibility of erosion 
due to abrasive wear and demonstrate that this 
hypothesis may account also quantitatively for the field observations. 
We will conclude the discussion of our model by deriving some theoretical 
predictions and compare them to those for thermal erosion.

{\em Description of the Laghetto channel} --
The channel, informally named Laghetto, is located on the upper southern 
flank of Mount Etna volcano at an elevation of 2560 m a.s.l.\ (see Fig.\ 1).
It is roughly linear, of rectangular-shaped cross section and  is incised 
on a lava slope gently inclined  (5$^{\circ}$--10$^{\circ}$). 
Its axis is oriented E-W. It is 220 m long and has a width between 
8 and 16 m. Maximum depth is  6 m, the floor of the channel is covered by lava.
The banks consist of lava flows emitted between July 25th and 30th. 
They are formed of up to 11 thin lava layers, each about 0.2 to 0.3 m thick 
(cf.\ Fig.\ 2). The eroded volume is about  220 m $\times$ 12 m $\times$ 4 m  
$\approx$ 10,000 m$^3$. Note that this volume is underestimated since part of 
the channel is filled with the eroding lava. 

{\em Description of the eruptive event} --
The eruption started on July 19th forming a crater in the area of Piano 
del Lago. In the first week, the eruption (closely observed by the authors) 
gave rise only to phreatomagmatic explosions and lava fountains. 
From July 25th for the 5 following days lava was erupted from the vent. 
The relatively low lava output (probably $ < 5$ m$^3$/s) was not constant 
but intermittent thus originating a number of thin and short 
(average length $< 500$ m) lava flows. 

In the evening of July 30th we witnessed an abrupt change in the lava output. 
A huge lava flow (10 to 20 m high) was emitted. The lava  flowed 
(in laminar motion) for about 12 hours reaching the area of Rifugio Sapienza, 
2000 m down valley. Near the vent the lava flowed on a bedrock formed by 
the still hot and soft thin lava layers of the preceding days. 
In these hours the bedrock was eroded and the following morning 
the Laghetto channel was there. After this last flow, lava emission 
at the Laghetto vent ceased. Therefore, the channel was not buried 
under subsequent lavas. It is this fortunate circumstance that allowed 
us to study the phenomenon and measure the channel before it was partially 
covered by sand and ash fall of the flank eruption of October 2002. 

{\em Thermal erosion hypothesis} --
Let us assume that the channel was thermally eroded. 
This implies that the lava flow had to supply (at least) the energy
$\Delta {\mathcal E}$
to melt the material originally contained in the channel, i.e., the 
volume $W_v\approx$ \mbox{$10^4$ m$^3$}
\begin{equation}
 \Delta {\mathcal E}\ =\ \rho\ W_v
                         \cdot (  \ L\ +\ c(T_m-T_{\rm subst}))\ \ .
\end{equation} 
Here $c= 1.4\cdot 10^3$ J/(kg K)  is the specific heat of the lava
substrate (calculated for the specific lava composition 
using MELTS~\cite{15.2}), $T_m\approx 1050^{\circ}$C is a lower 
limit for the onset of melting, and
$T_{\rm subst}\sim 900^{\circ}$C denotes the likely substrate temperature. 
With a lava density of $\rho=2800$ kg/m$^3$ and a latent heat of 
$L=0.5 \cdot 5.0\cdot 10^5$ J/kg for an assumed 50\% cristallinity
(using typical values for basaltic lavas~\cite{8,9,14.1}), 
about $1.3\times 10^{13}$ Joule were required. 

The eroding lava flow had a total 
length $s\approx $ \mbox{$ 2000$ m}
(the distance between vent and the Rifugio Sapienza 
area) and an average height $h_{\rm fl}\sim 10$ m. 
Typically, Mount Etna lavas are rich in phenocrysts 
(i.e., they are erupted partially solidified)
and their temperature ($T_l\sim 1050-1100^{\circ}$C) 
is close to the solidus point. 

Assuming an excessive temperature difference \mbox{$T_l-T_m \sim 50$ K}, 
the lava flow (approx.\ 240,000 m$^3$) could provide at most 
$4.7 \times 10^{13}$ Joule of heat energy (during 12 hours).
%
%
Therefore, a significant part of the 
available heat energy had to be transferred extremely rapidly from the 
lava flow to  the substrate to thermally erode the channel.  
The corresponding heat current density 
(transferred energy/(time$\times$channel surface)) amounts to
$\sim 10^5$ W/m$^2$).
Even with a temperature gradient on the order of $\sim 100$ K/m 
this corresponds to a heat conductivity of about $\sim 10^3$ W/(m$\cdot$K). 

This is  rather unrealistic as can be seen by the following 
considerations. 
First, the huge heat transfer rate would have required  a heat conductivity 
exceeding by far that of noble metals (e.g., for silver one has 
$\kappa_{\mathrm{Ag}}\approx 4\times 10^2$ W/(m$\cdot$K)~\cite{AshMer1976}
while basaltic lavas have $\kappa\sim 1$ W/(m$\cdot$K)~\cite{8,9}). 
Second, the noticeable heat loss would have increased viscosity 
leading to rapid deceleration and eventual standstill of the flow. 
This is in contradiction with the field observations.
Finally, we have neglected heat loss, e.g., radiative heat
loss and the formation of chill layers~\cite{9}.
Thus, we conclude that thermal erosion 
cannot have played a dominant role in the channel formation.

{\em Mechanical erosion hypothesis} --
Friction and wear are ubiquitous phenomena in every-day life. Their 
investigation in physics and engineering sciences, however, is a 
relatively young field~\cite{4,5}. Mechanical erosion is one of the 
fundamental processes in geology, quantitatively studied, e.g., 
in the context of bedrock~\cite{16} and glacial~\cite{6} erosion. While such 
studies focus on the dynamics, based on field data for many examples, 
the aim of the present discussion is merely to argue that the formation 
of the Laghetto channel does comply with the fundamental concepts of 
abrasive wear.

The idea of purely mechanical erosion is imposed by the fact that 
strata in the channel banks are clearly distinguished (cf.\ Fig 2), 
indicating that the material was removed by ``cutting'' rather than by melting.
The question is which mechanisms have acted and how they can be 
described quantitatively.

We propose the following intuitive picture.
The erosive lava was a thick (up to 20 m) autoclastic flow, i.e., a 
dense  mixture of crystals and clasts, metric in size, suspended in a 
melt matrix. The mesh of clasts got dragged along the bed such that
the irregular clast surfaces in contact with the bed could plough out grooves. 
These are the characteristics of abrasive and erosive wear 
processes~\cite{17}. 

It needs to be emphasized that abrasion and wear are rather complex phenomena
and therefore hard to model at a `microscopic' level (that is, by identifying
a dominant mechanism and deducing, e.g., wear rates 
from dynamical equations and material properties, in particular
if the process parameters cannot be defined as well as in systematic 
lab experiments). For this reason, phenomenological models that are
well-established by empirical studies and describe 
whole classes of wear processes are of considerable importance.
One of the simplest quantitative models to describe wear phenomena is 
Archard's law~\cite{4}.  It establishes a phenomenological relation between 
the wear volume $W_v$, the normal load $F$, and the sliding distance $s$:
\begin{equation}
      \frac{W_v}{s}\ =\ k\cdot \frac{F}{H} \ \ .
\label{Archard}
\end{equation}
Here,  $H$ denotes the hardness of the worn material proportional to 
the yield strength $\tau$ ($H\approx 3\tau$). Further, $k$ is the wear 
coefficient, i.e., a dimensionless constant  which depends on the wear 
mechanism. Abrasive wear by sliding glaciers is described by similar 
models~\cite{6}. 
From the observations we can estimate $W_v$, $s$, $F$, and $H$. 
For a realistic model, $k$ should be 0.1 or smaller. 
The wear volume is $W_v\approx  10,000$ m$^3$. 
The sliding distance is the total length of the flow ($s\approx  2000$ m) 
and the load $F$ is the weight of the abrading lava along the length 
of the channel. With an average height of the flow of 10 m we find  
$F\approx  26,000$ m$^3 \times  2800$ kg/m$^3  \times  9.8$ m/s$^2$ 
\mbox{$\approx  7.3  \times   10^8$ N.} 

It is more difficult to estimate the hardness $H$ 
of the substrate since its yield strength is an exponential function 
of temperature. Note that these lava layers had been erupted only days 
or hours before. Their outer skin was black, but they were still 
incandescent inside. For cold lavas yield strength values are $\sim 10^7$ Pa. 
According to Ref.~\cite{18}  for a temperature of 1000 $^{\circ}$C  
yield strength is larger than 10$^4$ Pa. For a likely temperature 
of 900 $^{\circ}$C we interpolate values between 10$^5$ to 10$^6$ Pa. 

By inserting these numbers into Eq.\ (\ref{Archard}) we obtain  
$k \approx  10^{-3} \ldots 10^{-2}$. 
This wear coefficient lies well within the typical range for abrasive
and erosive wear processes ($k\sim 10^{-5}\ldots 10^{-1}$) measured
in tribology~\cite{17}, 
confirming the intuitive picture for the erosive action of the lava flow. 

Note that our estimate of $k$ and its relation to the wear mechanism
is rather robust against variations
due to uncertainties in the observations while thermal erosion
remains excluded by several orders of magnitude. 
%

In view of the strong support that the experimental observations provide
for the idea of mechanical erosion, we may invert the argument:
we may ask what theoretical predictions can be obtained from 
assuming this kind of 
erosive mechanism, and compare them with the results for thermal erosion.
To this end, we rewrite  Archard's law Eq.\ (\ref{Archard}) in terms
of characteristic quantities of the erosive process. We have already
mentioned the sliding distance $s$, 
the height of the flow $h_{\rm fl}$, 
the density of the lava $\rho$, the hardness $H$ of the substrate,
and the constant $k$ which characterizes the wear mechanism.
Further we introduce the 
width $w_{\rm fl}$ of the flow  
(the latter is equal to the channel width $w_{\rm ch}$), the depth $d_{\rm ch}$ 
and the length $l_{\rm ch}$ of the channel. With 
$W_v=l_{\rm ch} w_{\rm ch}  d_{\rm ch}$ and 
$F=\rho g l_{\rm ch} w_{\rm fl} h_{\rm fl} \cos{\alpha}$ (where $\alpha$ is the
angle of the slope; for simplicity we will approximate $\cos{\alpha}\approx 1$
for moderate slopes) we obtain the relation
\begin{equation}
        d_{\rm ch}\ =\ \frac{k \rho g}{H}\  h_{\rm fl}\ s\ \ .
\end{equation}
This equation can be further transformed by assuming that the flow rate
per unit width $Q$ was constant for the duration $t_{\rm fl}$ of the flow.
The result is
\begin{equation}
        V_{\rm mech}\ =\ \frac{k \rho g}{H}\  Q\ \ ,
\label{mech}
\end{equation}
where $V_{\rm mech}$ denotes the velocity of mechanical erosion 
$V_{\rm mech}=\frac{\mathrm d}{{\mathrm d} t_{\rm fl}} d_{\rm ch}$. 
This relation can
be compared to Kerr's key result for the melting velocity 
$V_{\rm thermal}$ due to laminar flows  \cite{9}
\begin{equation}
        V_{\rm thermal}\ \propto\ 
             \left(\frac{U}{h_{\rm fl,thermal}}\right)^{1/3}
      \ \ ,
\label{Kerr}
\end{equation}
here $U$ denotes the surface velocity of the flow, 
and we have identified $d_{\rm ch}\equiv h_{\rm fl}$ for thermal erosion. 
While Eq.\ (\ref{Kerr}) implies, e.g.,~\cite{9}
\begin{equation}
   V_{\rm thermal}\ \propto\  Q^{1/9}\ \ ,\ \ \
   h_{\rm fl,thermal}\ \propto\  Q^{1/3}
\label{propkerr}
\end{equation}
our simple theory of mechanical erosion predicts
\begin{equation} V_{\rm mech}\ \propto Q\ \propto\ h_{\rm fl,mech}\ \ .
\label{propmech}
\end{equation}
Moreover, the substrate yield strength depends exponentially
on the substrate
temperature: $H=a \exp{(- b T_{\rm subst})}$
(with material constants $a$, $b > 0$) and thus 
determines the temperature dependence
of the ratio $V_{\rm mech}/Q$ in Eq.\ (\ref{mech}) while for thermal erosion
the ratio $V_{\rm thermal}/Q^{1/3}$ depends 
on $T_{\rm subst}$ via a rational function (the Stefan number, 
cf.\ Ref.~\cite{9}).
 
Thus, our hypothesis leads to concrete predictions which 
are substantially different from those for thermal erosion, even without
precise knowledge regarding the `microscopic details' of the erosion
mechanism. 
An important consequence of Eqs.~(\ref{propkerr}) and (\ref{propmech}) 
is that, with increasing flow rate $Q$, the velocity of mechanical erosion 
$V_{\rm mech}$ grows much faster than that of thermal erosion. That is,
for typical parameters of basaltic lava flows~\cite{9,18} one may obtain
velocities of mechanical erosion that are one or two orders of magnitude
larger than those for thermal erosion due to laminar flows 
with the same flow rate. This fact may help to explain the observation
of lava-channel formation even in the absence of very high
eruption temperatures and long-lived, 
low-viscosity flows~(cf.\ Ref.~\cite{14.1}).

{\em Conclusions} --
The analysis presented  here focused mainly on explaining the physical 
processes at the origin of the Laghetto channel. It may serve as a 
basic approach to mechanical erosion by lava flows. 
Further field studies and a better understanding of the erosive
processes at the 'microscopic level' will help to refine the model.
It is important to note that the observation of
 the channel formation described here represents a singular event. 
Nevertheless, similar features indicating mechanical erosion
can be found in historical and
prehistorical lava fields on Mount Etna and elsewhere~\cite{14.1}.  
Finally we mention that  
new aspects in channel formation processes on Earth may provide 
useful hints also for the assessment of erosion mechanisms in 
planetary environments such as lunar sinuous rilles and 
venusian canali~\cite{1,3,8}. 
\\
{\em Acknowledgments -- }
The authors would like to thank  R.\ Cristofolini and G.\ Ori 
for stimulating discussions.
J.S.\ is supported by a Heisenberg fellowship of the 
German Research Foundation.
\newpage
Fig.\ 1: Aerial  view of the Laghetto eruptive vent. 
The channel can be seen to the left of the pyroclastic cone. 
As indicated the erosional feature (head of the channel)
starts immediately at the emission vent.
Its length (220 m) gives the scale of the picture. 
In the upper part, there is the lava field produced by the eruption of 2001. 
\\[25mm]
Fig.\ 2: The right bank of the channel. In the foreground  
the bottom of the channel can be seen. It is covered by the eroding 
lava flow. The incision of the channel renders visible the layered 
structure of the substratum. 
It is important to note that these layers erupted few days or 
hours before were still rather soft and plastic. 
This is displayed by the uppermost layer folded inside the 
channel just after its incision.
\\[25mm]
Fig.\ 3: Cartoon showing  the mechanical erosion of the channel
         (not in scale). 
a) Near the vent (less than 500 m), the preflow terrain consisted 
of thin layers of hot lava. b)
While moving down the slope the autoclastic lava flow carved out the 
soft parts of the substrate. c) Situation after the flow had stopped. 
The channel floor is partially covered by lava of the eroding flow.
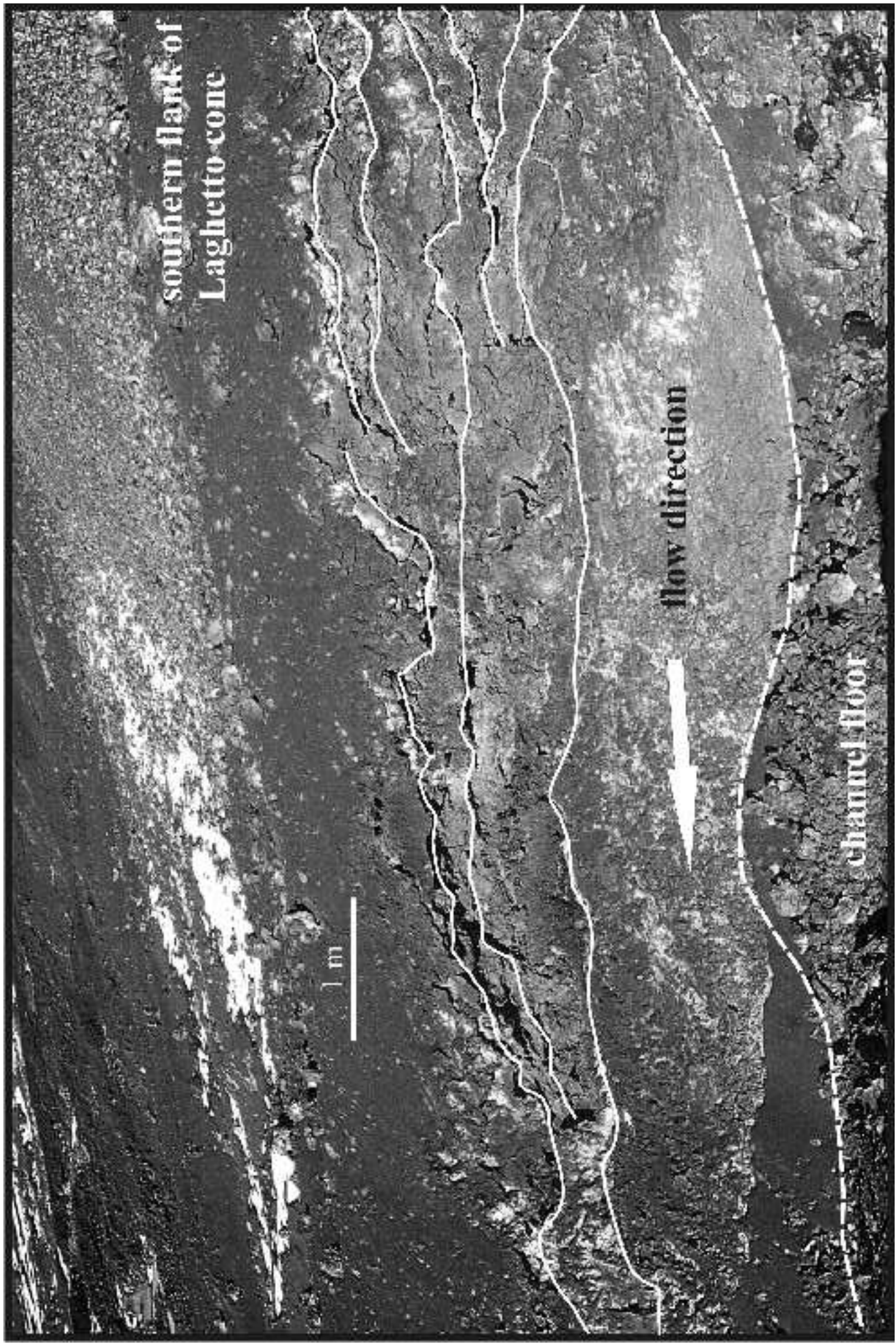
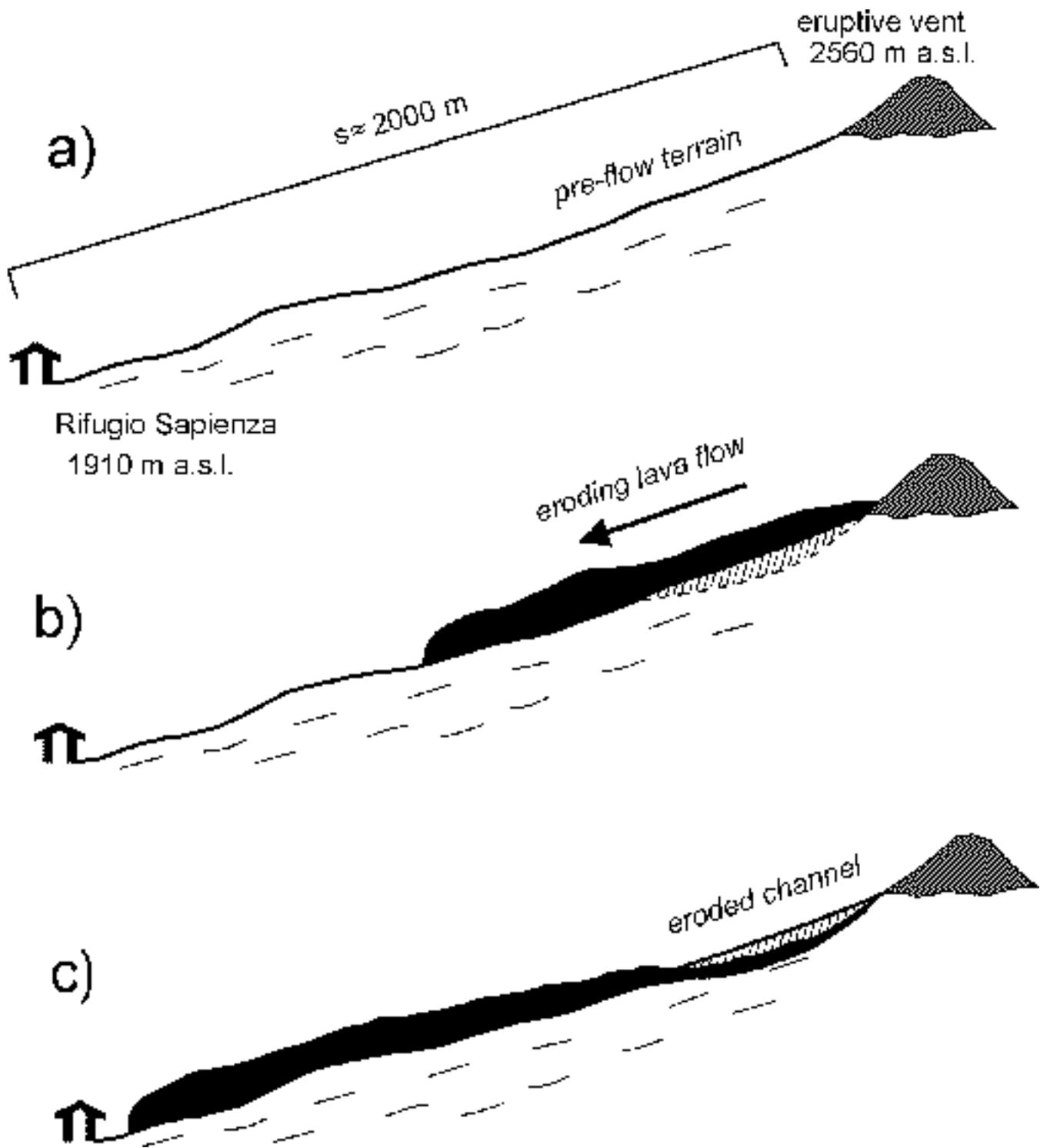
\end{document}